\documentstyle[12pt]{article}
\def \F{\phi}

\def\NP{{\it Nucl. Phys.\ }}

\def\PL{{\it Phys. Lett.\ }}
\def\PR{{\it Phys. Rev.\ }}
\def\PRL{{\it Phys. Rev. Lett.\ }}

\def\b{\beta}
\def\a{\alpha}

\def\s{\sigma}

\def\half{{1\over 2}}
\def\d{\dagger}
\def\be{\begin{equation}}
\def\eq{\end{equation}}
\def\Tr{{\rm Tr}}

\def\qd{{\rm QCD}_{2}}
\def\qq{{\rm QCD}_{4}}
\def\qt{{\rm QCD}_{3}}
\def\gt{{\tilde{g}}}
\def\r{{\rho}}

\begin{document}
\begin{flushright}
OUTP-9524P\\
hep-ph/9506456\\
\end{flushright}
\vspace{20mm}
\begin{center}
{\LARGE  Glueballs from $1+1$-Dimensional Gauge \\
Theories  with Transverse Degrees of Freedom \\ }
\vspace{30mm}
{\bf F. Antonuccio and S. Dalley}\\
\vspace{5mm}
{\em Department of Physics, Theoretical Physics\\
1 Keble Road, Oxford OX1 3NP, U.K.}\\
\end{center}
\vspace{30mm}

\abstract
We study  $1+1$-dimensional $SU(N)$ gauge theories with adjoint scalar
matter representations, based on a dimensional truncation of
$2+1$ and $3+1$-dimensional pure QCD, which approximate
the dynamics of transversely polarized gluons.
The glueballs  are investigated non-perturbatively using
light-front quantisation, detailed spectra and
wavefunctions being obtained for the large-$N$ limit.
In general there is some qualitative agreement of the spectra
with lattice Monte
Carlo data from the higher dimensional  QCD.
{}From the light-front wavefunctions we calculate (polarized)
structure functions and interpret the gluon
and spin content of glueballs.  We discuss the phase structure of the
reduced theories in relation to matrix models for relativistic
non-critical strings.

\newpage
\baselineskip .25in
\section{Introduction}

While there is little doubt that Quantum Chromodynamics \cite{qcd}
provides a correct description of most high energy processes inside
hadrons, our ability to determine its predictions for low energy
properties, such as  spectrum and wavefunctions,
is however much weaker. Of critical interest are the hadrons
with excited gluonic modes, glueballs, and hybrid mesons, whose
existence is  currently amenable to
concerted experimental investigation for perhaps the first time.
QCD has often been studied in $1+1$
dimensions, especially in the generalization to an infinite
number of colours $N \to \infty$, beginning with 't Hooft \cite{hoof}.
The reasons for doing this are both kinematical and dynamical.
Phase space in two dimensions is smaller and therefore simpler. Also many of
the characteristic features of hadrons, which are known or
expected to arise from highly quantum dynamical effects in
$3+1$-dimensional QCD ($\qq$), materialize as classical or mildly quantum
properties in $\qd$. Since there is no angular momentum
in one space dimension, conspicuously absent, however,
are any features of $\qq$ which rely on transverse degrees of freedom.
In particular there are no gluons at all. Is it therefore possible to
embellish $\qd$ so as to include some (discrete) remnant of
transverse fields without unduly complicating the theory?
In this paper we will study in light-front formalism
the glueball boundstates of a $1+1$-dimensional
gauge theory based upon a (classical)
dimensional reduction of  pure $\qq$, a model
of gluons which retains all polarizations of these particles but restricts
their 3-momenta to the form ${\bf k} = (k,0,0)$.
The result  is $\qd$ coupled to  two mutually interacting
adjoint scalars, corresponding to the two
possible physical polarizations of (massless) gluons
in four dimensions.
Despite the rather crude approximation of degrees of freedom,
our analysis in general can yield qualitative  agreement with spectra
obtained by lattice Monte Carlo simulations of pure $\qq$.
Having said this, our aim here is not to reproduce the numbers of
higher dimensional gauge theory exactly, since the reduced theory is
not well-suited to this task, but to treat the longitudinal and some
of the transverse dynamics precisely.
In this respect
the non-abelian reduced gauge model in $1+1$-dimensions seems to capture much
of the physics of the higher dimensional theory.

In the late seventies a remarkable attempt was made by Bardeen {\em et. al}
\cite{bill}
to compute the glueball spectrum directly from a
strong coupling expansion of large-$N$ light-front Hamiltonian
lattice gauge theory.\footnote{Review of more recent approaches to
solving four-dimensional light-front gauge theories with further
references can be found for
example in ref.\cite{revbrod}.} However, the  expansion scheme ultimately
left  an  effective potential for transverse motion
which had to be determined empirically.
Our calculation, based upon the dimensional reduction mentioned
above,
is formally similar, but obviously does not
include transverse kinematics, which would require knowledge of this
effective potential. In order to partially overcome the effects of
this
omission it is expedient to allow certainly one, and possibly two,
adjustable parameters
in the reduced theory, which are allowed by $1+1$-dimensional gauge
invariance
as we shall see.
The analysis we perform
is a natural extension of more recent work on the spectrum of
adjoint matter in $1+1$-dimensional large-$N$ gauge theories \cite{dk}.
The simpler dimensionally reduced model of $\qt$ at large-$N$
has been studied in ref.\cite{bdk1},
questions relevant to the Hagedorn behaviour (of adjoint
fermions) in refs.\cite{kut}, and the heavy  particle potential
in ref.\cite{tzani}. In this paper we
will first perform in more detail the broad spectrum analysis of
refs.\cite{dk,bdk1} in order to elucidate the
implications for glueballs in $\qt$, for which there is some
preliminary
lattice data \cite{mike}.

Our computations are performed in
the large-$N$ limit, which still however allows the number of gluons to
fluctuate, unlike the case for quarks.
This is done both for practical
reasons, in order to keep the  Hilbert space small enough to be
manipulated on a workstation when studied numerically, and also possibly for
reasons of principle in order to avoid  subtleties of
vacuum structure. The approach we use is analogous to the
original one of 't Hooft \cite{hoof} i.e. naive light-front
quantisation in light-front gauge.
While this is not in general good enough for studying vacuum
properties, such as the value of condensates,
it appears to be sufficient for the
excitation spectrum.\footnote{For a careful treatment of these
questions
see for example \cite{lenz,pauli}.}
To solve the boundstate problem numerically we employ discretised light-front
quantisation \cite{thorn,dlcq} conveniently implemented in terms of a
Lanczos algorithm, described in an appendix.
We check the numerical results analytically in a phase space
wavefunction (PSW) approximation.
The solutions yield a mass spectrum and  also light-front wavefunctions
from which virtually any structure function of interest can be
extracted.

The organisation of the paper is as follows. Section 2
introduces the light-front dimensionally reduced model.
New numerical and analytic solutions of
the boundstate equations for the reduced model of $\qt$
are described in section 3 and compared with the available lattice data.
Following the physical lessons learned from this, a similar analysis
is applied to the reduced model of $\qq$ in section 4;
the glueball
boundstate spectrum is compared with lattice
data and the structure functions are investigated. In section
5 we briefly look at the phase structure of the reduced model in more detail,
particularly in relation to matrix models of string theory which
have been studied in a similar light-front formalism \cite{dk2,ad}.
Finally,  section 6 summarises the physical conclusions we
draw from this work about non-perturbative QCD. Descriptions of the
numerical procedures  are as far as possible collected in the appendix.
An extension of the results to mesons and hybrids is treated in
\cite{meson}.

\section{Dimensional Reduction.}

We start from $SU(N)$ Yang-Mills theory in $D$ dimensions
\begin{eqnarray}
S & = & -{1\over 4 \gt^2} \int d^D x \ \Tr F_{\mu\nu} F^{\mu\nu} \ , \\
F_{\mu \nu} & = & \partial_{\mu} A_{\nu} - \partial_{\nu} A_{\mu}
+{\rm i} [A_{\mu}, A_{\nu}] \ ,
\end{eqnarray}
and choose light-front co-ordinates
$x^{\pm} = x_{\mp} =  (x^0 \pm x^{D-1})/\sqrt{2}$ and
light-front gauge $A_- = 0$. In this case
\begin{eqnarray}
S_{\rm lf} & = & {1 \over \gt^{2}} \int d^{D-2}x dx^+ dx^- \Tr \left[
-\partial_{+} A_\r \partial_{-} A^\r  + \half (\partial_{-} A_{+})^2 +
A_{+}J^{+} - {1 \over 4} F_{\r \s}F^{\r \s}
\right] \ ,  \label{fixed} \\
J^{+}_{ij} & = & -{\rm i}
[A_\r, \partial_{-} A^{\r}]_{ij} - \partial_{-} \partial_{\r}
A^{\r}_{ij} \ ; \ \ i,j = 1,\ldots,N \ ; \ \ \r,\s =
1, \ldots , D-2 \ .\label{current}
\end{eqnarray}
Repeated indices are summed.
The field $A_{+}$ is a constrained variable which does not propagate
in light-front time $x^+$, leaving only the transverse gluons $A_\r$ as
physical degrees of freedom. Much of the complexity of  gauge
theory is due to the linear term in the longitudinal momentum current
(\ref{current}), so one might as an approximation  study the
theory truncated to zero modes $\partial_{\r} A_{\mu} =0$ only. This is
the same as beginning with a $1+1$-dimensional adjoint gauge theory
\be
S_{R} = \int dx^+ dx^- \Tr\left[ -\half D_{\a} \F_\r D^{\a} \F^\r - {1\over
4g^2} F_{\a\b}F^{\a\b}  -  {tg^2 \over 4}  [\F_\r,\F_\s][\F^\r,\F^\s]
\right] \ \ ; \ \ \a , \b = \pm \ , \label{red}
\eq
where $g^2 = \gt^2/ \int d^{D-2}x$, $D_{\a} = \partial_{\a} + {\rm i}
[A_{\a}, .]$, $\F_\r = A_{\r}/g$, and $t=-1$.
$S_R$ inherits a subgroup of the Poincar\'{e}
and gauge symmetries of the $D$-dimensional theory. The residual
internal symmetries of $S_R$ resulting from gauge invariance
of $S$ are just the $1+1$-dimensional gauge symmetries
\be
A_{\a} \to UA_{\a}U^{\d} + {\rm i}(\partial_{\a}U)U^{\d} \ , \ \F_\r
 \to U\F_\r U^{\d} \ . \label{sym}
\eq
Due to the asymmetry of the dimensional reduction procedure, which
singles out one spatial direction, it is appropriate to allow
different longitudinal and transverse gauge couplings. Hence we take
$t$ in (\ref{red}) to be a free parameter, which is still consistent
with the reduced gauge invariance (\ref{sym}). In fact the potential
for
transverse fields $\F_{\r}$ in $S_R$ is the
only quartic gauge invariant term with the centre symmetry
$\F_\r \to \F_\r + const.$ . Higher order interactions are not
suppressed on dimensional grounds, but neither are they required for
renormalisability; in the interests of simplicity we will therefore
use the action $S_R$.
It acquires  the $SO(1,1)$ subgroup of $SO(D-1,1)$ Lorentz
symmetries of $S$ and also has discrete symmetries:
charge conjugation $C$ ($A_{ij}^{\mu} \to -A_{ji}^{\mu}$)
induces
\be
 C \ : \ \ \F_{ij} \to - \F_{ji}
\eq
on the $1+1$-scalars;
there also remains the one-dimensional parity
\be
P_1 \ : \ \ x^{D-1} \to -x^{D-1} \ ;
\eq
 finally there are the internal symmetries under exchanges of
the $\r$ index which, together with $P_1$, form a remnant of the
$SO(D-1)$ spatial rotations.

Analysing the boundstates of $S_R$ by light-front quantisation, and
taking the large-$N$ limit with $g^2 N$ fixed,
results in additional simplifications. The calculation
is similar to, but much richer than, the one performed by 't Hooft
for the quark--anti-quark pair \cite{hoof}. The
calculation
is richer since, unlike the fundamental representation for quarks,
 any number of adjoint particles may be involved in the
formation of  a boundstate.
With $x^+$ as canonical time, the $A_+$ field can be eliminated by its
constraint equation to yield canonical momenta:
\begin{eqnarray}
P^+ & =& -\int dx^- \Tr \left[  \partial_{-} \F_\r \partial_{-} \F^\r
\right] \ ;\label{pplus} \\
P^- & =& {g^2\over 2}  \int dx^- \Tr \left[ {t\over 2}
[\F_\r , \F_\s] [\F^\r , \F^\s]
-  J^{+} {1 \over \partial_{-}^{2}}J^{+}
\right] \ ; \label{pminus}
\end{eqnarray}
and the zero-charge condition $\int dx^- J^+ = 0$, where $J^+
= -{\rm i}[\F_\r, \partial_{-} \F^\r]$.
As in the t'Hooft model, the latter condition
restricts one to the space of singlet states under $x^-$-independent
gauge transformations, whose masses are gauge invariant.
Quantising on the
null line $x^+ = 0$,
\be
[\F_{\r ij} (x^-), \partial_{-} \F_{\s kl} (\tilde{x}^{-}) ]
= \half {\rm i} \delta_{\r \s} \delta (x^- - \tilde{x}^- ) \delta_{il}
\delta_{jk} \ ,
\eq
the boundstate problem reduces to the solution of the mass shell
condition,
\be
2 P^+ P^- | \Psi > = M^2 | \Psi > \ , \label{mass}
\eq
which, for a suitable  basis diagonal in $P^+$, is a
Schrodinger equation for the light-front hamiltonian $P^-$.
Such a basis is formed by the fourier  oscillator modes
$a_{\r}^{\d}$;
\be
\F_{\r ij}(x^+ =0) = {1 \over \sqrt{2\pi} } \int_{0}^{\infty} {dk^+ \over
\sqrt{2k^+}} \left( a_{\r ij}(k^+) {\rm e}^{-{\rm i}k^+ x^-}
+ a^{\d}_{\r ji} (k^+) {\rm e}^{{\rm i}k^+ x^-} \right) \ , \label{osc}
\eq
\be
[a_{\r ij}(k^+), a^{\d}_{\s lk}(\tilde{k}^+)] = \delta_{\r \s}
\delta (k^+ - \tilde{k}^+) \delta_{il} \delta_{jk}
\eq
These create the
light-front Fock space by action on a particle-free vacuum
$a_{\r}(k^+)| 0 > = 0$. We will often drop the superscript on $k^+$
for clarity and refer to it as momentum (it equals $k^0$ in the
traditional infinite momentum frame).
Due to the zero charge constraint  and the
large-$N$ limit the light-front wavefunctions one needs to consider are the
set of singlet states of the form
\be
|\Psi (P^{+}) > = \sum_{n=2}^{\infty} \int_{0}^{P^+} dk_{1}\ldots
dk_{n} \delta \left(\sum_{m=1}^{n}k_{m} -P^{+} \right)
{f^{\r_1 \ldots \r_n} (k_{1},\ldots,k_{n}) \over N^{n/2}}
 \Tr[a_{\r_1}^{\d}(k_{1}) \cdots
a_{\r_n}^{\d}(k_{n})] |0> \ .\label{wf}
\eq
For $SU(N)$, $\Tr[\F] =0$, and the one-gluon state is absent. The
glueballs
(\ref{wf})
may be pictured as a superposition of rings of flux created by a
closed chain of $n$ gluons
with momenta $k_{1} , \ldots , k_{n}$.
The states with more than one Trace, which are decoupled from these stable
flux rings when $1/N \rightarrow 0$,
correspond to multiple glueball combinations
and may be neglected. Explicit oscillator expressions may easily
be obtained for $P^\pm$ \cite{dk}.

The main objective of this paper is to solve (\ref{mass}) for
$D=3$ and $D=4$, and to explore the consequences. The elementary
interaction processes occuring at order $g^2$ are
shown in Figure 1,
the principal difference between $D=3$ and $4$ being the absence of the
four-gluon transverse-magnetic $t$-terms in the
former case (fig.1(c)).
We shall see that this difference has significant consequences
for the accuracy of valence gluon approximations and the phase
structure of the theory.
In the large-$N$ limit these proceseses must form Feynman
diagrams that  can be drawn on a cylinder, whose ends are the flux rings
(\ref{wf}), rather than a strip as in the case of quarks. The only
renormalisation which needs to be dealt with in the two-dimensional
theory concerns the self-energies shown in fig.1(a). There is no transverse
gauge invariance to prevent the generation of masses for the
fields $\F_\r$ and so a mass counter-term must be added to $S_R$ if
masslessness were to be maintained at the quantum level. In fact, as in
the t'Hooft model, the most divergent part of the self-energy is
cancelled by zero momentum $A_{+}$-exchange ($E$-term in fig.1(b))
when two or more gluons
are present. We elaborate these observations in the specific cases to
follow.

\section{Reduced $\qt$.}
\subsection{Boundstate Equations.}

First we reconsider the reduced model for $D=3$ \cite{dk,bdk1}.
There is only one
transverse gluon field, so we drop the $\r$ index and call
$f_{\r_1 \cdots \r_n} \equiv f_n$.
By substituting the oscillator expressions (\ref{osc}) into $P^{\pm}$,
normal ordering and
taking matrix elements, the mass shell condition (\ref{mass}) can be
conveniently summarised by an infinite set of coupled Bethe-Salpeter
integral equations for the
coefficients $f_n(x_1, \ldots x_n)$ of
$n$-gluon Fock states (cyclically symmetric in their
arguments);
\begin{eqnarray}
{M^2 \pi \over g^2 N} f_n  & = & {m^2\pi \over
g^2 N} {1 \over x_1} f_n (x_1, x_2, \ldots ,x_n) + {\pi \over
4\sqrt{x_1 x_2}} f_n (x_1, x_2, \ldots , x_n) \nonumber \\
& + & \int_{0}^{x_1 +x_2} {dy\over 4\sqrt{x_1 x_2 y (x_1 + x_2 -y)}}
\{ E[x_1,x_2,y] f_n (x_1 ,x_2,, \ldots,
x_n)  \nonumber \\
&& \nonumber\\
& +&
(D[x_1,x_2,y] - E[x_1,x_2,y]) f_n (y,x_1 +x_2 -y, \ldots, x_n) \}
\nonumber \\
&& \nonumber\\
& + &\int_{0}^{x_1} dy \int_{0}^{x_1-y}dz\ {F[x_1,-y,-z] + F[-z,-y,x_1]\over
4\sqrt{x_1 y z (x_1 - y -z)}}
 f_{n+2} (y,z,x_1-y-z,x_2,\ldots ,x_n)\nonumber \\
&& \nonumber\\
&+& {F[x_3,x_2,x_1] + F[x_1,x_2,x_3]\over
4\sqrt{x_1 x_2 x_3 (x_1 + x_2 + x_3)}}
f_{n-2}(x_1+x_2+x_3,x_4,\ldots,x_n)  \nonumber \\
&& \nonumber\\
& +& \ {\rm cyclic} \ {\rm permutations} \ {\rm of} \
(x_1,x_2,\ldots,x_n)
\ ,
\label{bs}
\end{eqnarray}
where the functions
\begin{eqnarray}
E[x_1,x_2,y] & = & {(x_1+y)(x_1 + 2x_2 -y) \over  (x_1 -y)^2} \\
D[x_1,x_2,y] & = & E[x_2, -y, -x_1] \\
F[x_3,x_2,x_1] & = & E[x_3, x_1, -x_2]
\end{eqnarray}
correspond to the elementary processes of fig.1.
In (\ref{bs}) we have switched to the Lorentz-invariant
momentum fractions $x_m = k_{m}^{+}/P^{+}$, which are the
Bjorken scaling variables; the two-dimensional light-front
gauge theory exhibits exact scaling.
The particle  $\F$ has logarithmic
and linearly divergent self-energy in $1+1$ dimensions as a result
of the first diagram in fig.1(a).
This diagram arises from normal ordering the $J.J$ term
in $P^-$ (\ref{pminus}) (the currents $J$ themselves are assumed already
normal ordered) and gives an induced mass \cite{bdk1}
\be
 \frac{1}{2} m^{2}_{\rm ind} = {g^2N \over 4\pi} \int_{0}^{\infty} {dp\over p}
+ {g^2 N \over \pi} \int_{0}^{k} dp {k \over (k-p)^2}
\eq
As in the t'Hooft model, the linearly divergent part of
$m^{2}_{\rm ind}$ cancels against zero-momentum $A_+$-exchange (fig.1(b))
between two particles and can be absorbed into the $E$-term in
(\ref{bs}) to make the integrals manifestly finite at the
Coulomb pole $x_1 =y$. To the rest of
$m^{2}_{\rm ind}$ we have added a bare $\F$-mass in order to leave a
finite
renormalised mass $m$ in (\ref{bs}).
In this paper we will mostly be interested in the massless limit $m \to 0$,
although no pathology would be incurred in choosing any
$m^2>0$.
In this way one can be sure that
whatever
constituent gluon masses appear to manifest themselves in the spectrum
are
dynamically generated. (We stress however that agreement with the
higher
dimensional lattice results can be improved if a constituent mass $m$
is put in by hand.)
Note that dimensional reduction of classical pure $\qt$ can be
achieved by compactification of one direction to very small radius,
but at the quantum level this  results in a diverging
Debye mass for Polyakov loops
\cite{doker}.
At small radius these loops can be identified with  $\F$-fields
and therefore our mass renormalisation prescription is entirely
different from the highly compact gauge theory, whose spectrum is quite
disimilar to the uncompactified one \cite{spain}.

The elementary processes (fig. 1) for adjoint particles represented
in (\ref{bs}) include the  Coulomb potential ($E$ integral), which is
linear in two dimensions; it would also
be present for fundamental representation
particles but with half the strength $g^2 \to  g^2/2$ since an adjoint
source has two flux lines attached to it rather than one.
In addition in (\ref{bs}) there is a $2 \to 2$
annihilation channel ($D$ integral) and pair
creation and annihilation  of gluons ($F$ integral), which are
not suppressed by the large-$N$ limit but for low-lying levels are
found to be dynamically suppressed  to a very high
degree \cite{dk}.

A survey of the broad features of the mass spectrum $M^2$ resulting
from large-$N$ adjoint scalars in two dimensions has been performed
in refs.\cite{bill,bdk1}. A massive spectrum of stable glueballs is organised
into approximate  valence gluon trajectories for low-lying levels, while at
higher energies pair creation dominates to give a complicated picture.
We have concentrated on the highly structured seven or so lowest states
resulting from (\ref{bs}) in
order to compare with  Monte Carlo results of the (unreduced)
lattice $\qt$.
Simple approximate analytic solutions may be found to the
equations (\ref{bs})  while numerical
solutions may be straightforwardly
 obtained  with the help of
{\em Mathematica}
and a workstation. By cutting the interval of allowed momentum
fractions $0<x<1$ into $x \in (1/K,3/K,5/K,\cdots)$ for some integer
$K$, the problem becomes  one of finite matrix diagonalisation
(Discretised
Light-Cone Quantisation)
\cite{bhp}, the continuum limit being achieved by extrapolating $K\to
\infty$.
We found it useful to employ a Lanczos
algorithm in
computing and diagonalising the mass matrix $M^2$ in this way.
A Hilbert space of dimension $\sim O(10^3)$ can be comfortably
handled. Details of our numerical work are given in the appendix.
In addition to measuring the masses $M$ from extrapolation of
finite $K$ calculations to the continuum limit $K =\infty$, we also
used the (normalised) light-front wavefunctions $|\Psi >$ to compute
structure functions of these mass eigenstates at fixed
cut-off.
In particular the quantities
\begin{eqnarray}
g(x) & = & <\Psi| a_{ij}^{\dagger}(x) a_{ij}(x) |\Psi >   \\
& = & \sum_{n=2}^{\infty} \int_{0}^{1} dx_{1}\ldots dx_{n} \delta
\left(\sum_{m=1}^{n} x_m -1\right)
\sum_{m=1}^{n} \delta ( x_m - x) |f_n|^2 \ , \nonumber \\
<n> & = & \int_{0}^{1} dx \  g(x) \ , \\
1 & =& \int_{0}^{1} dx \ x g(x) \ \ {\rm (momentum \ sum  \ rule)} \ ,
\end{eqnarray}
give the the number of gluons $g(x)$
with momentum fraction between $x$ and $x+dx$ in a boundstate and
the average number  of  gluons $<n>$.
These quantities are helpful in trying to
match the $1+1$-dimensional eigentates forming represenations of
discrete reduced symmetries with the representations $J$ of the full
spatial rotation group $SO(2)$ in $2+1$ dimensions.
In addition they suggest appropriate analytic approximations.

\subsection{Solutions.}

The  numerical solutions for the reduced $D=3$ theory are given in Table 1
and illustrated  on Figure 2 (some data for fixed $K$ are tabulated in
ref.\cite{bdk1}).
As in refs.\cite{dk,bdk1}
we note that a spectacularly accurate valence gluon approximation is
manifested in
the light glueballs described by (\ref{bs}).
The Coulomb potential, which includes the 2nd term in (\ref{bs}) as
well
as the $E$-integral, dominates since it is the
only positive definite and singular amplitude. Both the annihilation
channel ($D$) and pair production ($F$)
amplitudes take either sign with roughly
equal probability, leading to much cancelation. Physically they
are suppressed because the intermediate $A_+$ `particle' is
non-propagating in light-cone time $x^+$, so couples unfavourably to
transverse gluon pairs in the boundstate. The same reasoning may be
applied to the good valence approximation  for $\qd$ with massive
quarks
at finite $N$, which has been  studied in the same formalism
\cite{bhp}. Therefore the latter is  probably not a very good guide to
the accuracy of valence approximations for quarks in higher
dimensional QCD, where pair production can proceed via transverse
gluons.

In labeling the states with their $(P_1,C)$ quantum numbers,
a technical problem arises in determining $P_1$, which has the effect
$x^+ \leftrightarrow x^-$ on vectors. This symmetry is broken when the
momentum fractions $x$ are discretised, since this is equivalent to
(anti)periodic
conditions on $x^-$. Under $P_1$, momentum fractions of $n$ on-shell partons
transform as \cite{bhp}
\be
x_{m} \to {1 \over x_{m} \sum_{m'=1}^{n} {1\over x_{m'}}} ,
    \label{parity}
\eq
which in general is not of the form $({\rm integer})/K$; the exception
is $n=2$, in which case it has the effect $x_1 \leftrightarrow x_2$.
Also it is not obvious that eq(\ref{parity}) represents $P_1$ in the
interacting theory, although we found it to be at least an approximate
symmetry of the states exhibited in this paper.
\footnote{After
submission of this paper, similar work appeared for
$N=3$ \cite{N3},
where a fuller discussion of parity can be found. }
There are
various quantitative and qualitative
ways of assessing by inspection the effect of $P_1$ as given by
(\ref{parity})
on eigenfunctions at fixed
cut-off
$K$. The qualitative one
we found particularly useful was to note that in general
$P_1$ transforms  Fock states where one gluon carries most of the
longitudinal
momentum to Fock states where the momentum is shared evenly between
all the gluons.

A striking property of fig.2 is the almost
linear $M$ versus $<n>$ trajectory, which
continues much further than we have shown and which is approximately
repeated before soon becoming diluted by pair production effects at higher
mass. Such a trajectory  represents  `radial excitation' of the glueball
flux loop in the sense that the mass of a state is increased
by adding  gluons and their attendant flux lines to the ring.
The leading radial trajectory shows featureless structure functions
(Figure 3) peaked
at $1/<n>$, while the higher trajectory exhibits further oscillatory
behaviour analogous to  higher `angular momentum' states.
For the latter
there is an increased probability of finding an asymmetrical sharing of the
momentum between gluons in the glueball.

It is interesting to
note that $g(x)$ remains large for $x\sim 0$ in general, as a result
of the massless limit $m \to 0$. In particular
the approximately two-gluon states  are rather like cosine
wavefunctions, which was also found for the
low-lying
mesons in the 't Hooft model at small quark mass \cite{bhp}.
In the latter case the massless groundstate meson formed from
massless
 quark and anti-quark
had a constant wavefunction in phase space $f_2(x,1-x) = {\rm const.}$.
In the present case
the $CP_1 = ++$  groundstate wavefunction is approximately gluonium
\be
|\Psi> \approx \int_{0}^{1} dx \ {f_2 (x, 1-x)\over N}
\Tr [ a^{\d}(x) a^{\d} (1-x) ]|0>
\eq
The wavefunction $f_2(x,1-x) \equiv g(x)$
is not quite constant due to the
extra statistics-related contribution  $\pi / 4 \sqrt{x_1 x_2}$ to its mass
in (\ref{bs}), suppressing it at the
endpoints; this gives the glueball spectrum
its mass gap.  The gluonium approximation, where all $f_n$ are set to
zero except for $f_2$, is equivalent to the problem of scalar
fundamental representation matter \cite{fund}, with $g^2 \to 2g^2$. The
endpoint behaviour is $f_2 \sim x^{\a}$ for $x
\to 0$
where
\be
\a \cot{(\pi(\half-\a ))} =  {m^2 \over 2Ng^2} \ , \label{endp}
\eq
which is important for determining the form of $1/K$ corrections
(see appendix). Note that for $m \to 0$ one has  $f_2(x) \to {\rm const} >
0$
as $x\to 0$.
By arguments analogous
to those for quarks at finite $N$ \cite{bhp},
we expect this endpoint behaviour to
hold true beyond the gluonium approximation.

On the leading radial trajectory  the wavefunctions are well-described
by constant ones in phase space for most momentum fractions $x$. Only near
$x\sim 0$, corresponding to large $x^-$-separations,
 do interactions become appreciable.
Therefore one can obtain an
estimate for those masses by evaluating $M^2$ in a sub-basis of
phase space wavefunctions (PSW) defined by
\be
\psi_n : f_n (x_1, \ldots x_n) = {\rm const}. \ , \ f_{m} = 0 \ {\rm
for} \  m \neq n \ .
\eq
A  straightforward calculation using (\ref{bs}) shows that $M^2$ is
diagonal;
\begin{eqnarray}
 M^2 | \psi_n > = {n(n-1)\pi g^2 N \over 4} |\psi_n> \\
g(x) = n(n-1)(1-x)^{n-2} \ ,
\end{eqnarray}
in fair agreement with the numerical data (table.1). Only the 2nd term
on the r.h.s. of eq.(\ref{bs}),
$\pi / 4\sqrt{x_1 x_2}$, which is left over from
partial cancellation of the processes involving radiation of
zero-momentum $A_+$ (figs.1(a)(b)), contributes to the
eigenvalue; it is the principal value of the $E$ integral.
Since $M \sim n$ asymptotically, one
could regard  it as the  contribution to a constituent
 gluon mass
due to longitudinal interactions, or more appropriately as the groundstate
energy stored in
the flux string between each pair of neighbouring gluons on the flux
ring.
Practically speaking,
the PSW approximation
works well because the lowest radial trajectory starts quite a bit
lower than the complex higher trajectories.

\subsection{Comparison with $\qt$.}

Recent lattice Monte Carlo data for $SU(3)$ glueball masses extrapolated to
the continuum limit in three
dimensions
\cite{mike} is given in table 1 for comparison. Finite lattice spacing
errors are significantly reduced in three dimensions, so these results
can
be considered quite reliable.
The classification is $J^{PC}$, where
parity $P$
in two space dimensions is taken to be reflection about {\em one} axis (e.g.
$x^1 \to x^1$, $x^2 \to -x^2$).
Particles of non-zero angular momentum $J$ should be degenerate in $P$-doublets
$|J> \pm |-J>$.
Although we have no way of telling  the true strength of breaking of $SO(2)$
rotational  symmetry by the dimensionally reduced theory, which
reduces it to a $Z_2$ subgroup, we will attempt to match our states
with $J^{PC}$ labels. Firstly we take $P\equiv P_1$.
Under $180^{0}$ rotations $x^1 \to -x^1$ and  $x^2 \to -x^2$,
but although $\F$ is a $1+1$-dimensional scalar, it is the
component
of a $2+1$-dimensional vector, so we will assume this induces
$P_1$ and $\F \to -\F$. Since all the states we study are found to be
invariant
under
\begin{equation}
{\cal O}:\F_{ij} \to \F_{ji},
\end{equation}
 which reverses the ordering of gluons
around
a flux ring, this implies that  $CP_1 \equiv (-1)^{|J|\ {\rm mod2}}$, which
rules out half the possible states in three dimensions from the very
beginning. In fact it removes precisely one of the states from each
$P$-doublet. (We should point out however that states odd under ${\cal
O}$
can of course be constructed, but that they lie
consistently  much higher in the spectrum.)
One is still left with the problem of spin labelling
within the set of even and odd spins separately, because in general
there will be mixing within each set, i.e. spin 0 mixes with spin 2
etc..
We took the spin label  for states of given $(P_1,C)$ on a radial
trajectory to be the lowest which has not appeared with the same
$(P_1,C)$
on a lower energy trajectory.

Having truncated to zero transverse momentum, the best one could hope
for would be to reproduce the correct level ordering of the higher
dimensional theory.
The mass ratio scale gets expanded in the reduced theory, which one
could
ascribe to the level mixing  as a result of $SO(2)$-breaking.
It is tempting, in this gauge and in this large-$N$ light-front
quantisation scheme, to describe the data in terms of a
constituent gluon picture.
However, in at least one case this is in clear conflict
with the ($N=3$) lattice data: a $O^{--}$ state
could be interpreted as an orbital
excitation of a 3-gluon groundstate $1^{+-}$; on the lattice the
$O^{--}$ is the lower state since it couples to the combination
$U_{P} - U_{P}^{\d}$ of one Wilson plaquette $U_{P}$, while $1^{+-}$
requires
 a longer lattice Wilson loop. In fact table 1 shows that the  ($N=\infty$)
dimensionally reduced spectra in the $C=-1$ sector predict
lighter spin 1 glueballs, in disagreement with the 3D lattice.
The ordering in the $C=+1$ sector is quite good by comparison.
Spin 1's being lower than spin 2's  may be due to the mixing
of even spins and odd spins  amongst
themselves since spin  1 is pushed down by
mixing with spin 3 etc.
Due to the dominant valence content of wavefunctions, it is easy to
assess
the effects of a non-zero mass $m$ for gluons. Although this would act
to shrink the mass ratio scale
and push up the spin 1's, it
could not produce a lower-lying $O^{--}$.

Lattice results also exist for $SU(2)$
when $C=+1$ \cite{mike}, although
 states with $C=-1$  do not appear  since Wilson loops
are  unoriented.
We also therefore tried to assess the error in extrapolation
to $N=\infty$ by fitting to the form $M_{\infty} =
M_N -{\rm const.}/N^2$ for example.
The variation is actually quite small
and we do not think
that finite $N$ corrections are the major source of error, relatively
speaking, but that the discrepancies we have found are probably
largely  of a kinematic nature due to the severity of
the truncation and hence the smallness of
the residual $Z_2$ symmetry. Despite some agreement of the data, our
$P$ and especially $J$ assignments must be considered as heuristic
since mixing is inevitable. In many ways the reduced model of $\qq$,
which is the case of physical interest,
is less ambiguous.

\section{Reduced $\qq$.}

\subsection{Boundstate Equations.}

We now consider the reduced model for $D=4$, for which there are two
transverse polarizations.
Since all particles are restricted to move in one space dimension it
is especially  convenient to work in the helicity basis (c.f.eq(\ref{osc}))
\be
a^{\d}_{\pm} = {a^{\d}_{1} \pm {\rm i} a^{\d}_{2} \over \sqrt{2}} \ .
\eq
Then
\be
h = \int_{0}^{\infty} dk \ \{a_{+ij}^{\d}(k)a_{+ij}(k) - a_{-ij}^{\d}(k)
a_{-ij}(k) \}
\eq
measures the helicity of a state. The glueball
mass eigenstates  will  have definite helicity, since it is a
symmetry of $S_R$, and fall into
degenerate pairs of opposite $h$.
In general of course, these will not exhibit the further degeneracies
required to make up $SO(3)$ representations, a penalty of throwing away
all transverse momenta.

In the helicity basis the light-front momentum and energy
(\ref{pplus}),
(\ref{pminus})
read
\be
P^+ = \int_{0}^{\infty}dk \  k \{a_{+ij}^{\d}(k)a_{+ij}(k) + a_{-ij}^{\d}(k)
a_{-ij}(k)\} \ ,
\eq
\begin{eqnarray}
P^- & = & {m_{\rm ind}^{2} \over 2} \int_{0}^{\infty}
{dk \over k}\{a_{+ij}^{\d}(k)a_{+ij}(k) + a_{-ij}^{\d}(k)
a_{-ij}(k)\} \nonumber \\
&&+{g^2 \over 8 \pi } \int_{0}^{\infty}
{dk_{1}dk_{2}dk_{3}dk_{4}\over \sqrt{k_1 k_2 k_3 k_4}}
\delta(k_1 + k_2 -k_3 -k_4)\{
(D[k_1, k_2, k_3] + t)
a_{+ij}^{\d}(k_3) a_{-jk}^{\d}(k_4)a_{-il}(k_1) a_{+lk}(k_2)
\nonumber \\
&&+ (D[k_1, k_2, k_3] -E[k_1,k_2,k_3] - 2t)a_{+il}^{\d}(k_3) a_{-lk}^{\d}(k_4)
a_{+ij}(k_1) a_{-jk}(k_2)\nonumber \\
&& +
(t - E[k_1,k_2,k_3]) a_{+ij}^{\d}(k_3) a_{+jk}^{\d}(k_4)a_{+il}(k_1)
a_{+lk}(k_2) \} \nonumber \\
&& + \delta(k_1 + k_2 + k_3 - k_4)\{(F[k_1,k_2,k_3] - t)[
a_{+ij}^{\d}(k_1) a_{-jk}^{\d}(k_2) a_{+kl}^{\d}(k_3) a_{+il}(k_4) \nonumber \\
&&+a_{+ij}^{\d}(k_3) a_{-jk}^{\d}(k_2) a_{+kl}^{\d}(k_1) a_{+il}(k_4)
+a_{+il}^{\d}(k_4) a_{+ij}(k_1) a_{-jk}(k_2) a_{+kl}(k_3) \nonumber \\
&&+a_{+il}^{\d}(k_4) a_{+ij}(k_3) a_{-jk}(k_2) a_{+kl}(k_1)] \nonumber \\
&&+ (F[k_1,k_2,k_3] + t)[
a_{-ij}^{\d}(k_3) a_{-jk}^{\d}(k_2) a_{+kl}^{\d}(k_1) a_{-il}(k_4) \nonumber \\
&&+ a_{+ij}^{\d}(k_1) a_{-jk}^{\d}(k_2) a_{-kl}^{\d}(k_3) a_{-il}(k_4)
+a_{+il}^{\d}(k_4) a_{+ij}(k_3) a_{+jk}(k_2) a_{-kl}(k_1)\nonumber \\
&&+a_{+il}^{\d}(k_4) a_{-ij}(k_1) a_{+jk}(k_2) a_{+kl}(k_3)]\}
\nonumber \\
&&+ \ ( a_+ \leftrightarrow a_- ) \ .
\end{eqnarray}
The same set of elementary processes  occur (fig.1), with helicity
conserved at each vertex.
A set of integral equations similar to (\ref{bs}) may be derived; in
the following the polarization index is now taken as $\r = \pm$,
with the properties $\delta_{+-} = \delta_{+}^{\ +} = 1$,
$\delta_{++}=0$,
etc..
\begin{eqnarray}
{M^2 \pi \over g^2 N}f_{\r_1 \cdots \r_n}&  = & {m^2\pi \over
g^2 N} {1 \over x_1} f_{\r_1 \cdots \r_n}
(x_1, x_2, \ldots ,x_n) +
{\pi \over
4\sqrt{x_1 x_2}} f_{\r_1 \cdots \r_n}  (x_1, x_2, \ldots , x_n)
\nonumber \\ &&
\nonumber \\ &+ & \int_{0}^{x_1 +x_2} {dy\over 4\sqrt{x_1 x_2 y (x_1 + x_2
-y)}}
\{ E[x_1,x_2,y] f_{\r_1 \cdots \r_n}  (x_1 ,x_2,, \ldots,
x_n) \nonumber \\ && \nonumber \\
&+ &
(D[x_1,x_2,y] + t) \delta_{\r_1 \r_2} f^{\r}_{\ \r \r_3 \cdots \r_n}
(y,x_1 +x_2 -y, \ldots, x_n) \nonumber \\ && \nonumber \\
&+ &(4t\delta^{\r_1}_{\ \r_2} -3t - E[x_1,x_2,y])
f_{\r_1 \cdots \r_n}   (y,x_1 +x_2 -y, \ldots, x_n) \}
\nonumber \\ && \nonumber \\
& +& \int_{0}^{x_1} dy \int_{0}^{x_1-y}{dz\over 4\sqrt{x_1 y z (x_1 - y
-z)}}
\nonumber \\ && \nonumber \\ &&
\{(F[x_1,-y,-z] -t\ {\rm sgn}(\r_1){\rm sgn}(\r))
f_{\r_1 \   \r \r_2 \cdots \r_n}^{\ \ \r}
(y,z,x_1-y-z,x_2,\ldots ,x_n) \nonumber \\ && \nonumber \\
&+&
(F[-z,-y,x_1] +t\
{\rm sgn}(\r_1){\rm sgn}(\r))
 f^{\r}_{ \ \r \r_1 \cdots \r_n}
(y,z,x_1-y-z,x_2,\ldots ,x_n)\} \nonumber \\ &&
\nonumber \\
&+& {1 \over 4\sqrt{x_1 x_2 x_3 (x_1 + x_2 + x_3)}}
\{(F[x_3,x_2,x_1]+t\  {\rm sgn}(\r_1) {\rm sgn}(\r_2) ) \times
\nonumber \\
& & \nonumber \\ & & \hspace{3mm}
  \delta_{\r_2 \r_3}f_{\r_1 \r_4 \cdots \r_n}
(x_1+x_2+x_3,x_4,\ldots,x_n) \nonumber \\ && \nonumber \\
& +&
(F[x_1,x_2,x_3] +t\ {\rm sgn}(\r_2) {\rm sgn}(\r_3) )
\delta_{\r_1 \r_2} f_{\r_3 \cdots \r_n}
(x_1+x_2+x_3,x_4,\ldots,x_n)  \}\nonumber \\ &&
\nonumber \\
&  +& \ {\rm cyclic} \ {\rm permutations} \ {\rm of} \
(x_1,x_2,\ldots,x_n)\ .
\label{bs2}
\end{eqnarray}
The mass
renormalisation proceeds in exactly the same fashion --- the only new
ingredient is the normal ordering of the $t$-terms which gives a
further logarithmically divergent contribution to $m_{\rm ind}^2$ ---
and it is in the form (\ref{bs2}) that we discretise the momentum
fractions
$x$ for
numerical computation.

The variable
$t$ is considered as a free parameter, which we will adjust to obtain
a physically reasonable spectrum. In particular we find that it must
be positive in order for the groundstate glueball to be helicity
zero.\footnote{This result was also found for the similar
transverse plaquette
coupling in the light-front hamiltonian lattice gauge theory of
ref.\cite{bill}.}
Unlike in $\qt$, gluon pair production can now proceed directly
from a contact term fig.1(c). While it is therefore less suppressed,
for a range of $t$ around zero there is still a discernable valence
gluon content for light glueballs.
If $t$ is too large however, tachyons appear and
eventually there appears to be a phase transition; the phase structure
is discussed in more detail in section 5.
As in section 3, we  primarily study the  massless case
$m=0$. However
we will find once again that
there is an expansion of the mass ratio scale in the reduced theory
relative to that obtained from the higher dimensional lattice results,
which can be mitigated by including a constituent mass $m$ by hand.

{}From
the wavefunctions $\Psi$ (\ref{wf}) we also calculate both
unpolarized
\be
G(x) = {< \Psi| a_{+ij}^{\d} (x) a_{+ij}(x)+ a_{-ij}^{\d} (x)
a_{-ij}(x) \ |\Psi > \over  <\Psi | \Psi >}
\eq
and polarized
\be
\Delta G(x) = {< \Psi| a_{+ij}^{\d} (x) a_{+ij}(x) -  a_{-ij}^{\d} (x)
a_{-ij}(x) | \Psi > \over  <\Psi | \Psi >}
\eq
structure functions.
As a consequence of no transverse momenta, there is a trivially
satisfied helicity sum rule
\be
\int_{0}^{1} dx \ \Delta G(x) = h \ .
\eq
In higher dimensions these structure functions would also be
integrated over transverse momenta up to some scale $Q$, the evolution
with large
$Q$ being calculable within perturbative QCD once the  function
is specified at a given scale. In the present case $Q=0$
effectively, and we are performing a non-perturbative calculation of
(some of) the long-distance properties of boundstates in $\qq$ scarcely
known for any hadron, least of all glueballs.

\subsection{Solutions.}

Results for masses of the seven lightest glueball states at
$t=0.15$ ($m=0$) are given in Table 2 and illustrated on Figure 4
classified by $|h|^{P_1 C}$.
For massless gluons $m=0$ we are forced to choose a relatively small
$t$ in order to avoid tachyons, so it is the longitudinal interactions
which are still dominant.
As expected, the valence approximation is less accurate than for the
$D=3$ reduced model, but is nevertheless manifest. All the states
shown
are groundstates with respect to their valence gluon content and,
with the exception of $0^{-+}$,
exhibit
featureless unpolarized structure functions (Figure 5) peaked at $1/<n>$.
The $x \to 0$ behaviour is still governed by eq.(\ref{endp}).
The negative $P_1$-parity state  $0^{-+}$, like the $0^{--}$ for
$D=3$, has strong longitudinal
interactions, which is why it appears  relatively heavy.
We attribute the persistence of valence gluon dominance to the energy
required to add extra flux lines (one per gluon) to the ring, since the
study of purely longitudinal interactions in section 3 showed that each
contributes roughly $\sim g\sqrt{N\pi/8}$ to the mass.

Once more, it is possible to estimate the light spectrum
in the PSW approximation by setting
\be
f_{\r_1 \ldots \r_n}(x_1, \ldots, x_n) = {\rm const.} \label{fgw4}
\eq
for a given set of gluon helicities $\r_1 , \ldots, \r_n$.
This provides a sub-basis of wavefunctions $\psi_{\r_1 \ldots
\r_n}$  for the `radial' excitations of
the
groundstate in each $|h|^{C}$ sector ($P_1 = +$).
The integrals appearing in the Bethe-Salpeter equations
(\ref{bs2}) can be done exactly for (\ref{fgw4})
resulting in a finite $M^2$ matrix
for
finite maximum $n$. Note that $M^2$  is now no longer diagonal in each sector
due to the $t$-terms. Diagonalising in the sub-basis of PSW's
up to $n=6$  gives the estimates
shown as $M^2_{\rm PSW}$ in table 2. As an example we show the $M^2$
sub-matrix in the PSW basis
$\{\psi_{+-},\psi_{++--},\psi_{+-+-},\psi_{++},
                        \psi_{+++-} \}$:

\begin{eqnarray}
  \left( \begin{array}{ccccc}
           \frac{\pi^2}{2} (1-t) & 2 \pi \sqrt{6} t &
-4 \pi \sqrt{3} t & 0 & 0 \\
2 \pi \sqrt{6} t & 3 \pi^2 (1-t/2) & \frac{3 \pi^2}{\sqrt{2}} t & 0 & 0
\\
-4 \pi \sqrt{3} t & \frac{3 \pi^2}{\sqrt{2}} t & 3 \pi^2 (1-2 t)& 0 & 0

\\
0 & 0 & 0 &  \frac{\pi^2}{2} (1+t) & 0 \\
0 & 0 & 0 & 0 & 3 \pi^2
         \end{array}
      \right).
\end{eqnarray}

Polarized structure functions of $h=0$ glueballs are identically zero
while that of the $|h|=2$ state follows closely the unpolarized
function because the non-valence content is small. The spin structure of
the $|h| =1$ state is perhaps the most interesting since it is the
nearest thing to a `nucleon' in the spectrum. Its larger non-valence
content means that the unpolarized function is peaked at noticeably
smaller $x$ than the polarized one, gluon pairs, which are preferably
created at small $x$, having to be
produced with opposite helicity. In Figure 6 we plot the helicity
asymmetry $A= \Delta G /G$ for the $1^{+-}$ glueball.
The PSW approximation
would give $A(x) =1/3$. At small $x$ one finds $A(x) < 1/3$ so the
helicity tends to be more disordered, while at large $x$ one has $A(x)
> 1/3$. In real nucleons one expects $A \to 1 $ as $x \to 1$ \cite{bb}
(helicity
retention) i.e. a parton carrying most of the momentum also carries
most of the helicity. Although this question has presumably not been
asked of glueballs before, it does not seem to happen in the reduced model for
$t = 0.15$, at cut-off up to $K=13$ at least,
although the $x=1$ intercept of $A$ does increase with
$t$.
Perhaps this an indication
that the strength of transverse interactions we are forced
to consider at $m=0$ are too small
to accurately reflect the higher dimensional theory.

\subsection{Comparison with $\qq$.}

Lattice Monte Carlo simulations of the (quenched) glueball
spectrum have been relatively rare and are still crude for all but the
lightest state (see e.g. refs.\cite{mike2,chen,bali}).
We used the $SU(3)$ results of ref.\cite{bali} for comparison with the
large-$N$ reduced
model since they include a range of light states. However that
computation was performed at fixed lattice spacing ($\b = 6.4$
on a $32^4$ lattice) so one must bear in mind that significant
variations in mass ratios are possible when extrapolated to the
four-dimensional continuum limit; furthermore, no `radial' excitations were
computed. It is generally believed that $N=3$ is large in the sense
that there is little qualitative difference between $SU(2)$ and
$SU(3)$ spectra.
Lorentz invariance was restored
in the lattice results to the extent that a full $J^{PC}$
classification was possible, but this is clearly not the case in the
reduced model, where $h^{P_1 C}$ are the exact quantum numbers. As in
the $D=3$ reduced model, some of the components of given light spin $J$ are
pushed higher in spectrum,
although each light spin $J$ does seem to be represented by its largest
$|J_z|$ values (compare helicity $h$ with spin $J$ in Table 2). In
particular the reduced model does not seem to cope well with
states odd under ${\cal O}: \F_{ij} \to \F_{ji}$,
which are needed to make up
components with $C = -(-1)^{|J_z|\ {\rm mod}2}$ (e.g. $J_z=\pm 1$ for
$J^{PC} = 2^{++}$). The results of Table 2 also show
that mixing between given $h^{P_1 C}$  seems
to prevent the degeneracy with $h=2$ of an  $h=0$ component,
which mixes with excitations of the groundstate scalar glueball $0^{++}$.
Finally we note that $P_1$ alone seems to be closer to three dimensional
parity $P$, rather than $P_1$ combined with $\F_{\r} \to -\F_{\r}$,
since
the parity of the lightest spin 1 would be incorrect in the latter
case.

Assuming that the $h$ assignments of the reduced model
are a good guide to the $J$
assignments in four dimensions, the level ordering is in qualitative
agreement with the lattice result, although once again the mass ratio
scale is expanded. Our model (with $m=0$) prefers a heavy $0^{-+}$,
due to strong longitudinal interactions, which we have treated
exactly, while the four-dimensional lattice
would prefer it to be lighter. It is
quite possible that the full effects of
transverse interactions, of which we have
only retained a remnant, and transverse kinematics shrink the
mass ratio scale and make the $0^{-+}$ lighter. To illustrate
this possibilty one can include a constituent mass $m$ by hand,
which one might regard as generated by the transverse dynamics that
the reduced model neglects. In Table 2 we have also given the spectrum
at $K=12$
for a set of parameters $m^2=2g^2 N /\pi, t=0.5$. The non-zero gluon
mass allows us to choose a value for  $t$ comparable to the longitudinal
interaction strength. The  comparison  with the ($N=3$)
four-dimensional lattice data
shows  much closer agreement, although no amount of
fudging can cause the degeneracies necessary for restoration of
$SO(3)$ rotational symmetries; the transverse kinematics are clearly
needed
for this to happen. The main effect on the structure functions is to
suppress
small $x$ partons (c.f. eq.\ref{endp})),  the polarization
asymmetry $A$ also being modified slightly (fig.6).

\section{Phase Structure and Matrix Models.}

In principle there are two adjustable parameters in the reduced model
for $\qq$, the mass $m$ and quartic self-coupling $t$. In this section,
by deriving bounds on $M^2$ and quantifying $K$-dependence, we suggest
a phase diagram in the $(m^2,t)$ plane and note the
analogy with matrix models of non-critical string theory. As is well
known, the planar Feynman diagrams for the field theory of a
hermitian matrix $\F_{ij}$ with action
\be
S= \int d^d x \Tr \left[ (\partial \F)^2 - \mu \F^2 + {\lambda\over
4N}
\F^4 \right] \label{mm}
\eq
define a certain type of random surface embedded in $D$ dimensions
\cite{surf}, essentially a discretisation of Polyakov's continuum
path integral formulation \cite{poly}. Such theories have been analysed
for $d=2$ and $N=\infty$ in light-front formalism \cite{dk2,ad}
in very much the same way as
the two-dimensioanl adjoint gauge theories. The action (\ref{mm}) is
merely ungauged. The critical value of $\lambda/\mu$ beyond which
perturbation theory diverges --- individual normal-ordered planar diagrams are
UV finite --- manifests itself in light-front formalism (without zero
modes) as the divergence $M^2 \to -\infty$
in the spectrum. Such a transition\footnote{The
possibility of phase
transition in the adjoint gauge theory when contact self-interaction
terms are added was already suggested in refs.\cite{ks,dk}.}  occurs for
the reduced model of $\qq$ also, separating a parton phase
with $<n>$ finite from a string phase with $<n> =\infty$.

Firstly, one can prove that the spectrum is unbounded below if $m^2<0$ by
using the variational wavefunction
\be
\int_{0}^{1} dx \ \Tr[ a^{\d}_{+}(x) a^{\d}_{-}(1-x)] |0> \ .
\eq
Secondly, that the spectrum is unbounded from below if $t$ is
sufficiently large can be proven with the (discretised) variational state
\be
\Tr[a_{+}^{\d} (1/K) a_{-}^{\d} (1/K)a_{+}^{\d} (1/K) a_{-}^{\d} (1/K)
\ldots a_{+}^{\d} (1/K) a_{-}^{\d} (1/K)]|0> \ .
\eq
Therefore the phase diagram is probably as indicated in Figure 7. In
principle
it is possible that the point $A$ coincides with $m=t=0$ and that the
spectrum is unbounded below for all $t>0$ when $m=0$. The
$1/K$-extrapolation curve of the groundstate $M^2$ shown in the
appendix
certainly seems to be dropping rapidly. However, although we could not
prove it, our experience suggests that unboundedness is unlikely since
we could not find a variational wavefunction which proved that $M^2 =
-\infty$ for all $t>0$ at $m=0$ --- it is certain that $M^2$ is positive at
$t=m=0$ --- while the matrix model analysis
makes definite
predictions for the $K$ dependence \cite{ad}, which can be manifested at quite
small $K$, for the divergences of the string phase;
\be
<n> \sim K \ \ , \ \ M^2 \sim -K^2 \label{dep} \ .
\eq
Our measurements at $t=0.15$ showed no sign of a linear divergence of
$<n>$, rather it seemed to converge to $\sim 2.06$.  However, at $t=0.45$
for example we find the $K$-dependence of eq.(\ref{dep}),
suggesting that this point is already in the string phase.
The precise details at the transitional values of $t$ are difficult to
determine at the small values of $K$ with which we are forced to work
here, but it would be interesting to  know whether there exists a
non-critical string regime analogous to that identified in
refs.\cite{dk2,ad}.
Another question that we are unable to answer at this stage
concerns
 the
role of $k^+ = 0$ modes on the phase structure, since they probably
have to be taken into consideration in the string phase for a faithful
treatment of the original field theory (\ref{red}). Finally we note
that if our conclusion concerning boundedness were wrong for $m=0$,
one would be forced to
introduce a sufficiently large gluon mass $m$ for given $t$ in order
to stabilize the spectrum.

\section{Conclusions.}

Two-dimensional non-abelian
gauge theories have proved to be usefully tractable, especially in
light-front
formalism and the large-$N$ limit.
In this paper we have attempted to incorporate some of the transverse
degrees of freedom of pure gauge theory in a two-dimensional
context. In this way one can investigate the boundstate properties of
physical gluons  and spin while postponing many of the subtle problems of
renormalisation which  arise in higher dimensional light-front
formalism. Although our asymmetrical
truncation of
phase space may seem an arbitrary approximation, the highest penalty
paid appears to be the lack of Lorentz multiplet degeneracies in the
glueball spectrum. Comparison with the lattice Monte Carlo data for both three
and four dimensions showed nevertheless that the dimensionally reduced
model always produces some components of each light Lorentz multiplet, and
in approximately the same level order.
The rogue components which do not appear in the light spectrum are
states anti-symmetric under the
symmetry transformation $A_{ij} \to A_{ji}$ of the gauge potential;
this fact deserves a better understanding.
Although one cannot be sure of the significance of the mechanism in
higher dimensions, the two-dimensional  model simply generates
from the longitudinal Coulomb potential
what amounts to a constituent gluon mass structure for the spectrum.
This is due to the non-zero groundstate energy for the flux string
between neighboring gluons in the flux loop. It is weaker than a true
gluon mass and so cannot prevent the wavefunction from spreading in
$x$-space when the gluon mass $m=0$, although there is a mass
gap in
the spectrum. It was also
the origin of a
good valence approximation for the $D=4$ reduced model. Such an
approximation is
even more evident for the $D=3$ model in which case pair production
must proceed via a  non-propagating longitudinally polarized $A_+$
field. An improvement in
the comparison with the higher dimensional spectrum was noticed when an
explicit gluon mass $m$ is included by hand --- this might be
neccessary anyway to bound the spectrum from below --- perhaps
compensating for neglected transverse dynamics.
The light-front structure functions exhibited physically reasonable
behaviour in our chosen region of coupling constant space,
being quite close to those of a phase space approximation. More
interesting
will be their content at finite $N$, which will govern the decay and
formation of glueballs, and when quarks are included.

Although many qualitative features of higher dimensional QCD are
reproduced
by the reduced models, we are sceptical of its use as the basis for a
systematically improvable approximation to Lorentz-invariant quantities
at the quantitative level as
transverse
momenta are added. The limit of zero transverse momentum is a very
singular
one in the quantum theory \cite{doker}, which we have avoided by
reducing
at the classical level and treating subsequent undetermined couplings
phenomenologically. Of course we would be happy to be proved wrong in
this scepticism! Rather we consider the transverse lattice approach
advocated in ref.\cite{bill} to be the most promising for quantitative
work --- a study is currently in progress --- whose mathematical structure
is virtually isomorphic to the field theories studied
in this paper, although the adjoint fields have a different interpretation.
Nearly co-linear phenomena in which transverse momenta are small
compared to a typical scale, such as high-energy diffractive
scattering, may hold the best opportunity for applying the reduced
models quantitatively to a {\em bona fide} four-dimensional problem.
With this in mind, the  scattering amplitudes involving the reduced glueballs
and mesons \cite{meson} are presently under study.

\vspace{5mm}
\noindent Acknowledgements: We would like to thank I. Klebanov, I. Kogan,
J. Paton, and M. Teper for very valuable discussions; M. Teper for
making available to us his lattice results before publication;
M.Burkardt and B.van de Sande for pointing out confusions in an
earlier manuscript.
F.A. is supported by the Commonwealth Scholarship and Fellowship Plan
(British Council).
\newpage

\begin{center}
{\sf APPENDIX.}
\end{center}
\noindent {\bf The Lanczos Algorithm.}

The Lanczos algorithm \cite{lanc,hiller}
is an iteration method which may be used to
determine numerically the eigenvalues and corresponding eigenstates
of a finite linear operator. We may summarise the essential ideas
as follows: Given a symmetric eigenvalue problem
 $A \mbox{\boldmath $u$} = \lambda \mbox{\boldmath $u$}$,
we begin by choosing a normalised vector ${\mbox{\boldmath $u$}}_1$,
and setting the constant $b_1$ to zero. A sequence of constants
$a_n,b_n$ and orthonormal vectors ${\mbox{\boldmath $u$}}_n$
are then generated by the following recursion relations;
\begin{eqnarray*}
  {\mbox{\boldmath $v$}}_{n+1} & = & A {\mbox{\boldmath $u$}}_n
                            - b_n {\mbox{\boldmath $u$}}_{n-1} , \\
     a_n & = &  {\mbox{\boldmath $v$}}_{n+1} \cdot
                                  {\mbox{\boldmath $u$}}_n , \\
       {\mbox{\boldmath $v$}}_{n+1}^{'} & = &
   {\mbox{\boldmath $v$}}_{n+1} - a_n {\mbox{\boldmath $u$}}_n ,\\
    b_{n+1} & = & \sqrt{{\mbox{\boldmath $v$}}_{n+1}^{'} \cdot
                {\mbox{\boldmath $v$}}_{n+1}^{'}} , \\
  {\mbox{\boldmath $u$}}_{n+1} & = &
               {\mbox{\boldmath $v$}}_{n+1}^{'} /  b_{n+1} .
\end{eqnarray*}
It is easy to show that the matrix representation of the operator $A$
with respect to the orthonormal vectors $\{{\mbox{\boldmath $u$}}_n \}$
is tridiagonal, with $a_n$ and $b_n$ being the diagonal and codiagonal
entries respectively.
After a given number of iterations, the eigenvalues and eigenvectors
of this tridiagonal matrix approximate those of the original operator,
with a progressive improvement in precision as the iteration proceeds.
A particular feature of this method is that the extreme eigenvalues
are readily obtained to high precision after a comparatively small
number of iterations. This makes the algorithm especially useful if
one is only interested in studying the low-lying states.

In our investigations, the finite linear operators which need to be
diagonalised arise from the discretised versions of the
integral equations (\ref{bs})
and (\ref{bs2}),
 in which only
odd\footnote{Choosing odd multiples gives rise to a faster rate of
convergence towards the continuum limit since the endpoint regions of
the $x$-interval $(0,1)$ are better sampled.}
 momentum fractions $ x_i \in \{ 1/K, 3/K , 5/K , \ldots \} $
are allowed, coupled with the constraint $x_1 + \cdots + x_n = 1$.
This enables us to replace  the functions
$f_{\r_1 \ldots \r_n} (x_{1},\ldots,x_{n})$ with a finite number
of components
\[
    g_{\r_1 \ldots \r_n} (m_1, \cdots ,m_n) =
     \frac{1}{\sqrt{K^{n-1}}} f_{\r_1 \ldots \r_n} (\frac{m_1}{K},
                                          \cdots ,\frac{m_n}{K}),
\]
where the  odd integers $m_i$ satisfy $m_1+ \cdots + m_n = K$. Note
that this cutoff regulates both momentum and the number of particles.
Thus the linear action of the integral equation (\ref{bs2}) on the
space
 of functions
$f_{\r_1 \ldots \r_n}$
may be represented in the discrete case as a finite constant
matrix whose size is determined by the number of independent components
$g_{\r_1 \ldots \r_n}(m_1,\ldots,m_n)$. For example, if we set
$K=4$, and consider the zero helicity sector of the reduced $3+1$
theory, the only relevant components are $g_{+-}(1,3),g_{+-}(3,1),
g_{++--}(1,1,1,1)$, and $g_{+-+-}(1,1,1,1)$. The corresponding
$4 \times 4$ symmetric matrix representing the discretised
version of the linear action (\ref{bs2}) has the explicit form
\begin{eqnarray}
  \left( \begin{array}{cccc}
        \frac{2 \pi}{\sqrt{3}}+ \frac{16}{3} x - \frac{4}{3} t
                     +\frac{16}{3} &  -\frac{16}{3} - \frac{4}{3} t &
           \frac{4}{\sqrt{3}} t & -\frac{8}{\sqrt{6}} t \\
    -\frac{16}{3} - \frac{4}{3} t &
    \frac{2 \pi}{\sqrt{3}}+ \frac{16}{3} x - \frac{4}{3} t
       +\frac{16}{3} &
  \frac{4}{\sqrt{3}} t & -\frac{8}{\sqrt{6}} t \\
  \frac{4}{\sqrt{3}} t &  \frac{4}{\sqrt{3}} t &
  4\pi +16 x -4 t & \frac{8}{\sqrt{2}} t \\
   -\frac{8}{\sqrt{6}} t & -\frac{8}{\sqrt{6}} t &
    \frac{8}{\sqrt{2}}t &  4\pi +16 x -16 t
           \end{array}
      \right) ,
\end{eqnarray}
where $x=\pi m^2 / g^2 N$. Diagonalising the above matrix gives the
eigenvalues $M^2$ in units $ g^2 N/\pi$.

One strategy which may be used to estimate the continuum (i.e.
 $K \rightarrow \infty$) limit of the theory is to calculate
these
matrices explicitly for increasing values of $K$, and then to
extrapolate the resulting sequence of eigenvalues.
Unfortunately, this procedure is severely limited by the rapid
growth in the number of independent components
$g_{\r_1 \ldots \r_n}(m_1,\ldots,m_n)$ as $K$ is steadily increased,
making it sometimes difficult to provide reliable estimates of
the continuum behaviour of the theory. Approximation methods
are then necessary if further progress is to be made.

As in a previous investigation \cite{ad}, we used Sun and  Dec-alpha
workstations together with Mathematica to implement the Lanczos algorithm
outlined earlier --- this way is slow but easy to do.
This technique enabled the authors to investigate
the cases $K \leq 13$ in the reduced $3+1$ theory and $K \leq 22$ for $2+1$.
Typically, the algorithm
requires a judicious choice of initial vector to ensure that the
convergence of the low-lying eigenvalues is well behaved and rapid.
 Owing to the
valence-like structure of the low-lying states, a perfectly adequate
initial state consists of two or three valence gluons (depending
on whether the helicity is even or odd respectively).
It is also important that it have definite symmetry under the
operators which commute with the  light-cone hamiltonian since,
as the iteration proceeds, each of the new vectors
formed by the action of the light-cone hamiltonian will possess
the same quantum numbers as the initial vector. This helps restrict the size
of the vector space one needs to work with. We remark that
for $K=12$, the $0^{++}$ sector consists of approximately $900$
linearly independent vectors,
while for $K=13$, the size of the $1^{+-}$ sector is
a little over $1800$.

In the context of discretised light-cone quantisation,
exploiting the advantages offered by the  Lanczos algorithm
depends crucially on how one implements the method. Typically, a large
number of vectors are produced at each stage in the iteration,
and so, as the iteration proceeds, the light-cone hamiltonian
will repeatedly act on the same vectors. However, this action
is far from trivial, and so the procedure rapidly runs into
difficulties due to the enormous growth in the number
of vectors that need to be dealt with.
Therefore, it was necessary to introduce a system of
`pointers' which connect a set of existing vectors with its
image under the action of the hamiltonian. This conveniently
eliminates the need to perform the same calculations repeatedly,
and dramatically improves the viability of the method.

\noindent {\bf Extrapolation.}

A fool-proof method of extrapolating the results of discretised
light-cone
quantisation to the continuum limit is still lacking.
The extrapolation of finite-$K$ calculations to the continuum limit
$1/K=0$ has been discussed in refs.\cite{bhp}.
Errors due to discretisation of the integrals in equations such as
(\ref{bs}) begin at $O(1/K)$ due to the Coulomb pole. In addition,
endpoint errors lead to dependence upon $K^{-\a}$
(c.f. eq(\ref{endp})).
Thus for
the case $m=0$ one expects the corrections to form a series in $1/K$
and one can fit $n$ data points taken at different $K$ to a
polynomial (Richardson extrapolation)
\be
M^2 = a_0 + {a_1 \over K}+ {a_2 \over K^2} + \cdots + {a_{n-1} \over
K^{n-1}} \ ,
\eq
with $a_0$ the continuum limit. $a_{n-1}/K^{n-1}$ evaluated at the
lowest
value of $K$ employed gives an order of magnitude estimate for the
error in extrapolating . These are the errors shown in the tables,
which are probably underestimated.
We illustrate this procedure below with the masses from the $D=4$
reduced model (section 4) at $t=0.15$, since the extrapolation is far
from straightforward especially for the groundstate in this case.
Given the suppression of non-valence gluons, one can increase the maximum
value of $K$ attainable numerically by truncating the Fock space to a
few gluons (Tamm-Dancoff approximation), thus enabling a more reliable
extrapolation to $K=\infty$.
Figure 8 illustrates the polynomial fits for $K \leq 12,13$ for all
states shown in table 2, and also includes the fit for the groundstate
obtained from data for $K \leq 18$ by using the $2-,4-,6$-gluon sector
only (chain line).

\newpage
\begin{center}
FIGURE and TABLE CAPTIONS
\end{center}
\noindent
{\bf Table 1} - $M$ is the $K=\infty$ extrapolated mass in units of
$\sqrt{Ng^2/\pi}$ based upon exact diagonalisations up to $K=22$.
$<n>$ is calculated for illustration from the theory truncated to
the 2-,4-,6-gluon sector only for $K=20$, or
to the  3-,5- gluon sector for $K=19$. The heuristic $J$ assignments are
discussed
in text. The lattice $SU(3)$ masses $M_{SU(3)}$ have  groundstate $O^{++}$
normalised to $M$ and a $2\s$ statistical error estimate (ours).
\vspace{5mm}

\noindent
{\bf Table 2} - Exact $M^2$ in units of $Ng^2/\pi$ for various $K$ at
$m=0$
and $t=0.15$; the continuum limit;  $<n>$  at $K=12$;
$M^2_{m}=M^2$ at $K=12$ for $m^2= 2g^2N/\pi$ and $t=0.5$; $<n>$ for the
latter. Also
given are mass squared ratios $\tilde{M}^{2}$ to the groundstate.
Lattice $SU(3)$ mass squared ratios to the groundstate $\tilde{M}^2_{SU(3)}$
are taken from the calculation of
ref.\cite{bali} at fixed lattice spacing and are rounded to two figures here.
\vspace{5mm}

\noindent
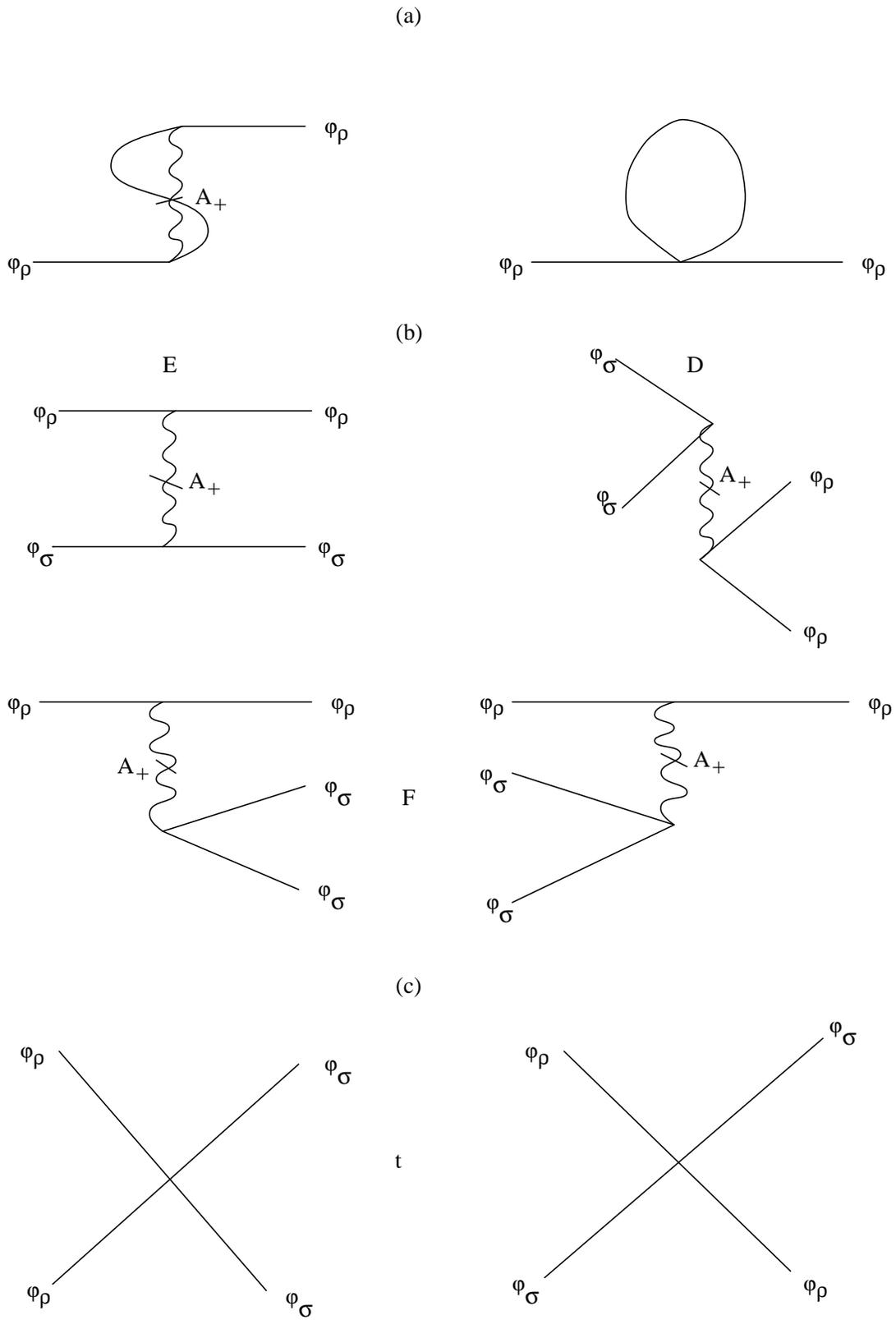
{\bf Figure 1} -- The elementary processes contributing at order $g^2$
in $S_R$: (a) self-energy; (b) longitudinal interactions; (c)
transverse interactions.
\vspace{5mm}

\noindent
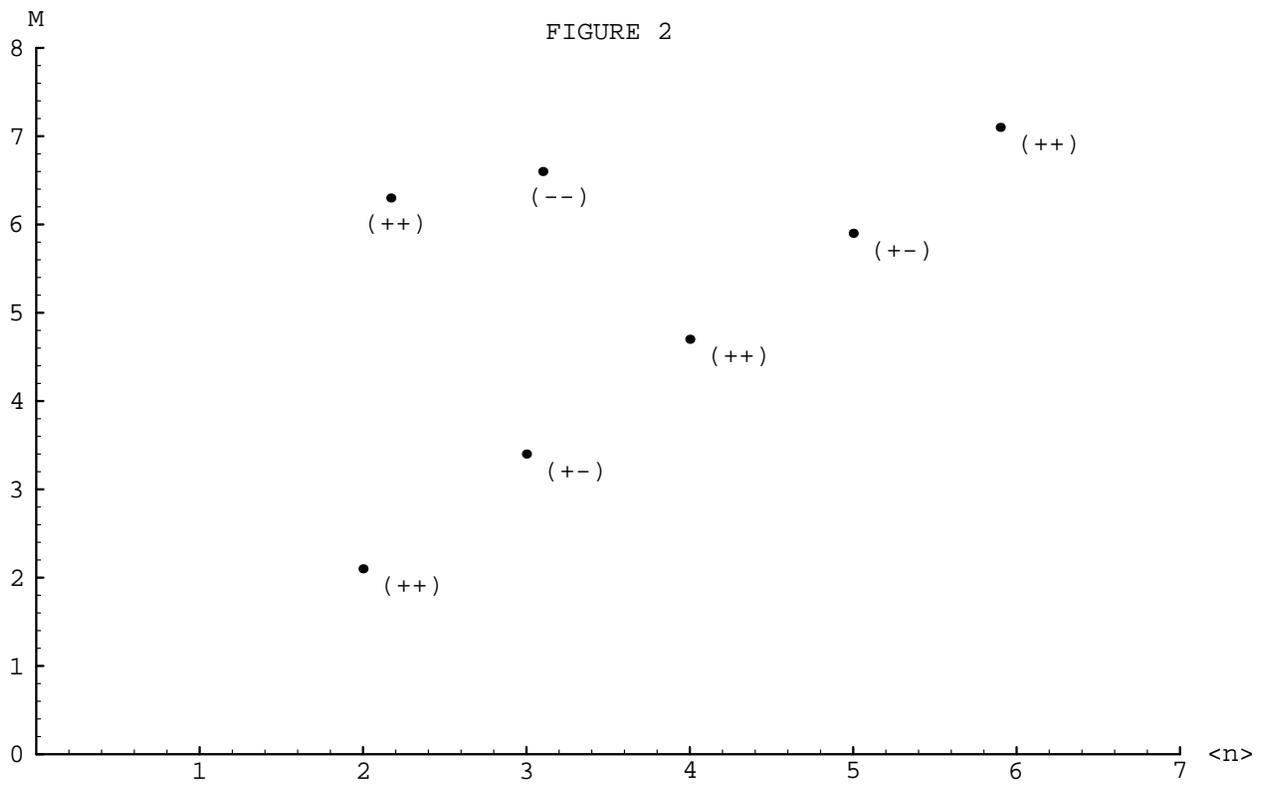
{\bf Figure 2} -- Mass spectrum, in units of $\sqrt{Ng^2/\pi}$,
extrapolated to the continuum limit $K = \infty$ and
classified by ${P_1, C}$.
\vspace{5mm}

\noindent
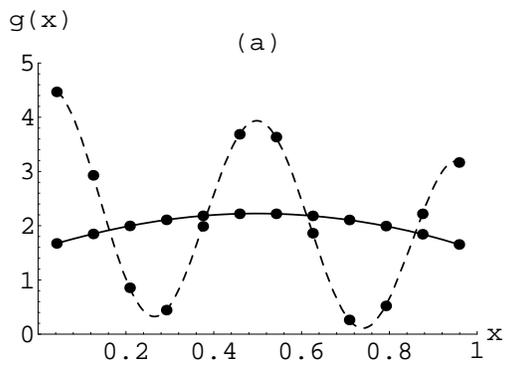
{\bf Figure 3} --
Gluon structure functions $g(x)$ of glueball mass
eigenfunctions. Full lines refer to states on the lower `radial'
trajectory in fig.2 and chain lines to the states on the upper trajectory.
 (a) $<n> \sim 2$ ; (b)  $<n> \sim 3$ ;(c) $<n> \sim 4$ ;
(d) $<n> \sim 5$.

\vspace{5mm}

\noindent
{\bf Figure 4} -- Mass squared spectrum in units of $Ng^2/\pi$ at
$m=0$ and $t=0.15$,
extrapolated to the continuum limit $K = \infty$ and
classified by $|h|^{P_1, C}$ (non-zero $|h|$ are doublets).
\vspace{5mm}

\noindent
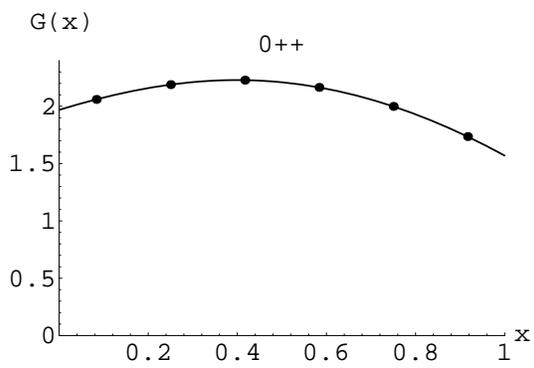
{\bf Figure 5} -- Gluon structure functions $G(x)$ for $m=0$ and
$t=0.15$
\vspace{5mm}

\noindent
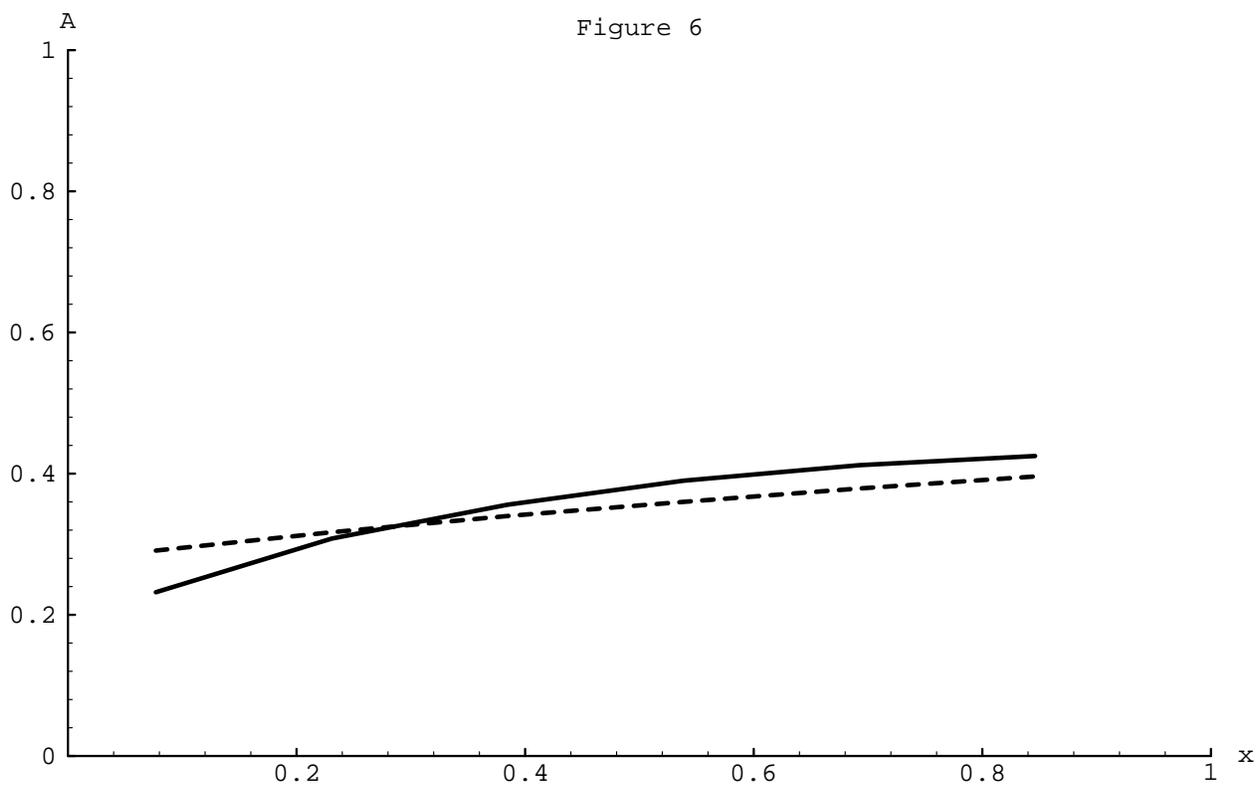
{\bf Figure 6} -- Polarization Asymmetry $A(x) = \Delta G/G$ of the
$1^{+-}$ glueball at $K=13$. Dotted
line
for $m=0$ and $t=0.15$, full line for $m^2=2g^2 N/\pi$ and $t=0.5$.
\vspace{5mm}

\noindent
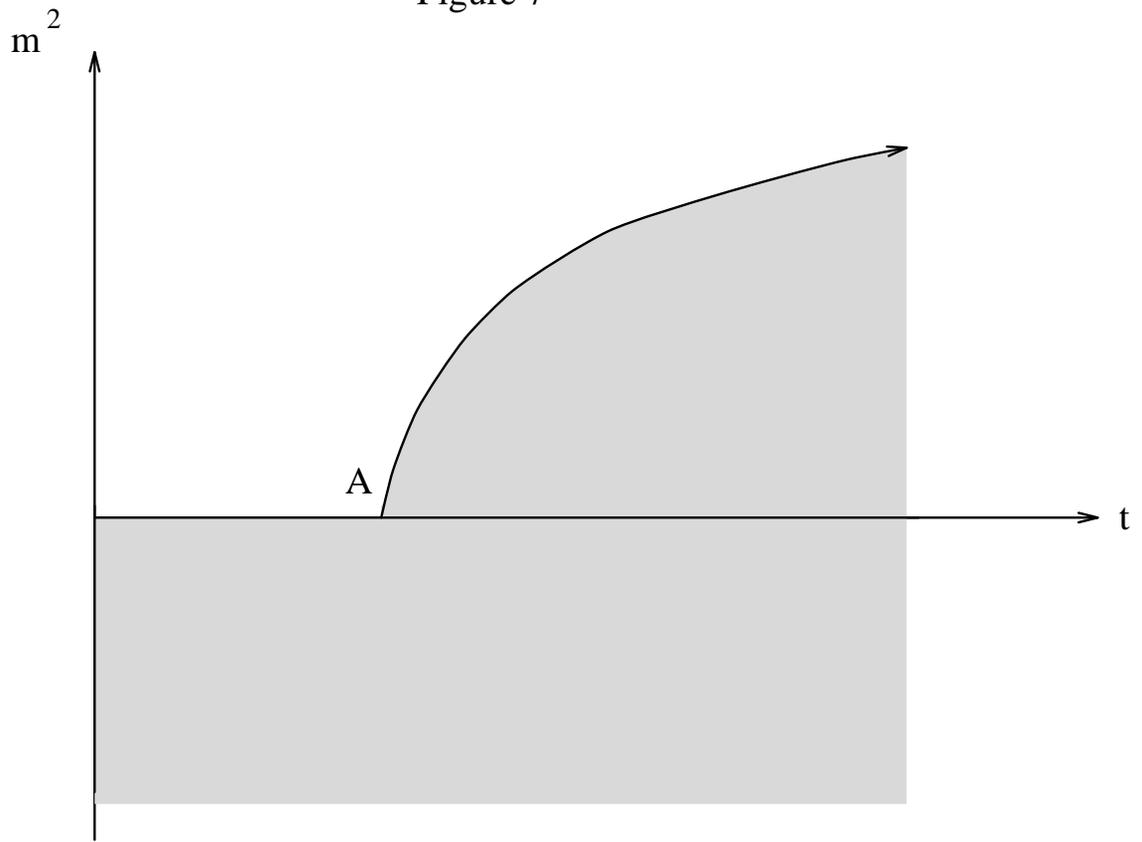
{\bf Figure 7} -- Suggested $(m^2,t)$ phase diagram. The shaded region
has
unbounded $M^2 \to -\infty$, while the theory is non-tachyonic in a
neighbourhood of the postive $m^2$-axis.
\vspace{5mm}

\noindent
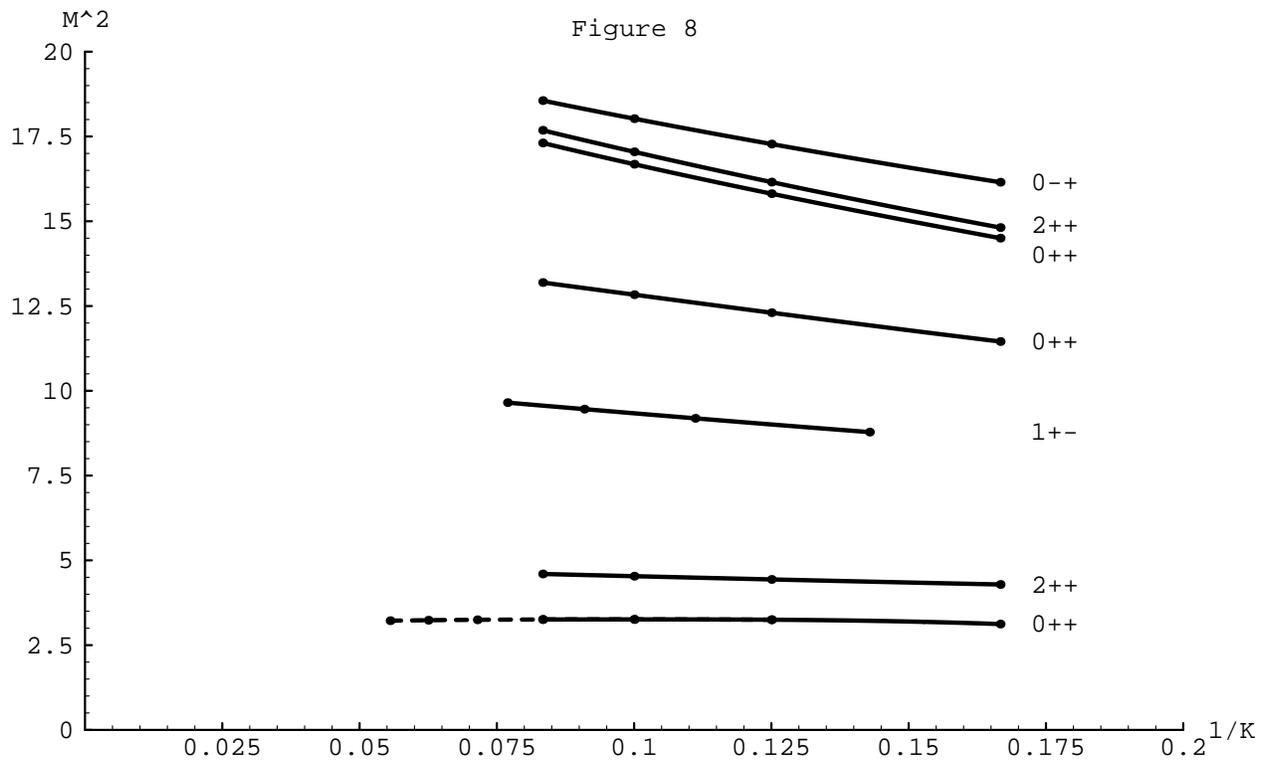
{\bf Figure 8} -- Polynomial fits to $M^2$ data (units of $g^2N/\pi$)
at $t=0.15$ and m=0.

\newpage

\noindent Table.1
\medskip
\medskip

\begin{tabular}{|c|c|c|c|c|c|c|c|} \hline
{\bf REDUCED} \\ \hline
$P_1 C (J)$   & $++(0)$ & $+-(1)$ & $++(0)$ & $+-(1)$
       & $++(2)$ & $--(0)$ & $++(0)$  \\ \hline
     M & 2.1(1) & 3.4(1) & 4.7(1) & 5.9(1) & 6.3(1) & 6.6(1) & 7.1(1) \\ \hline
     $<n>$ & 2.002 & 3.002 & 4.000 & 4.990 & 2.175 & 3.085 &
       5.902  \\ \hline
$M_{\rm PSW}$ & 2.2 & 3.8 & 5.4  & 7.0  & -- & -- & 8.6 \\ \hline
{\bf LATTICE}   \\ \hline
 $J^{PC}$     & $0^{++}$ & $0^{++}$ & $0^{--}$ & $2^{\pm +}$ &
$0^{--}$ & $0^{++}$ & \\ \hline
 $M_{SU(3)}$     & 2.1(1) & 3.2(1) & 3.2(1) & 3.5(3) & 4.0(2) & 4.0(2) &
\\ \hline
\end{tabular}

\medskip
\vspace{10mm}

\noindent Table.2
\medskip
\medskip

\begin{tabular}{|c|c|c|c|c|c|c|c|} \hline
{\bf REDUCED} \\ \hline
$|h|^{P_1 C}$   & $0^{++}$ & $2^{++}$ & $1^{+-}$ & $0^{++}$
       & $0^{++}$ & $2^{++}$ & $0^{-+}$  \\ \hline
 valence & $a_{+}a_{-}$ & $a_{\pm}a_{\pm}$ & $a_{\pm}a_{\pm}a_{\mp}$ &
$a_+a_-a_+a_-$ &     $a_+a_+a_-a_-$ & $a_{\pm}a_{\pm}a_{\pm}a_{\mp}$ &
$a_+a_-$ \\ \hline
 $K=    6$ & 3.12186 & 4.28929 & $8.78276^{1)}$ & 11.453 &
    14.502 & 14.814 & 16.1509 \\ \hline
     $K=8$ & 3.24943 & 4.43805 & $9.18900^{2)}$ & 12.303 &
    15.813 & 16.1555 & 17.2769 \\ \hline
   $K=10$ & 3.25875 & 4.53335 & $9.46014^{3)}$ & 12.835 &
    16.686 & 17.0475 & 18.0253 \\ \hline
  $K=12$ & 3.25605 & 4.59893 & $9.65198^{4)}$ & 13.193 &
    17.309 & 17.6828 & 18.5584 \\ \hline
  $K=\infty$ & 3.1(3) & 4.9(1) & 10.7(2) & 14.9(4) &
    21.0(4)& 21.3(9) & 21.7(3) \\ \hline
$<n>$ & 2.062 & 2.002 & $3.031^{4)}$ & 4.011 & 4.021 & 4.025 &
       2.134 \\ \hline
$M^2_{\rm PSW}$ & 3.1 & 5.7 & 13.3 & 17.0 & 27.9  & 28.6  & - \\ \hline
$M^2_{m}$ & 14.56 & 18.21 & $37.29^{4)}$ & - & -  & - & 35.98\\ \hline
$<n>$ & 2.042 & 2.005 & $3.027^{4)}$ & - & - & -& 2.022  \\ \hline
$\tilde{M}^2_{m}$ & 1 & 1.25  &  2.56 & - & -  & -  & 2.47 \\ \hline
{\bf LATTICE}  \\ \hline
$J^{PC}$ & $0^{++}$ & $2^{++}$ &  $1^{+-}$ &  && &$0^{-+}$ \\ \hline
$\tilde{M}^2_{SU(3)}$ & 1 & 1.5(1) & 1.9(2) & &  & &1.5(2) \\ \hline
\end{tabular}

${}^{1)}K=7,{}^{2)}K=9,{}^{3)}K=11,{}^{4)}K=13$

\vfil
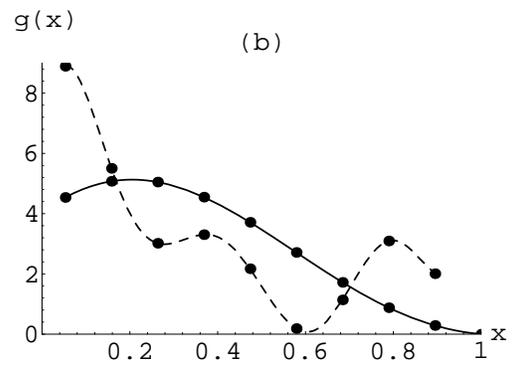
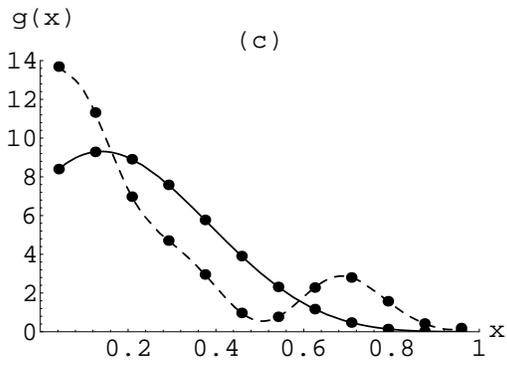
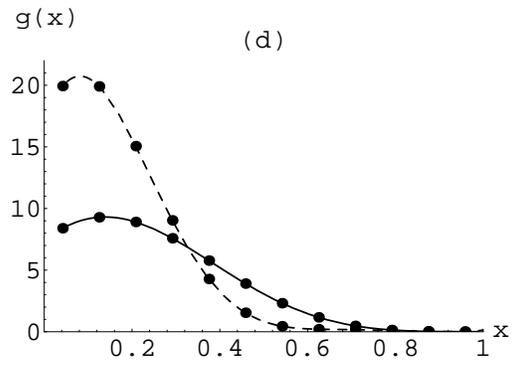
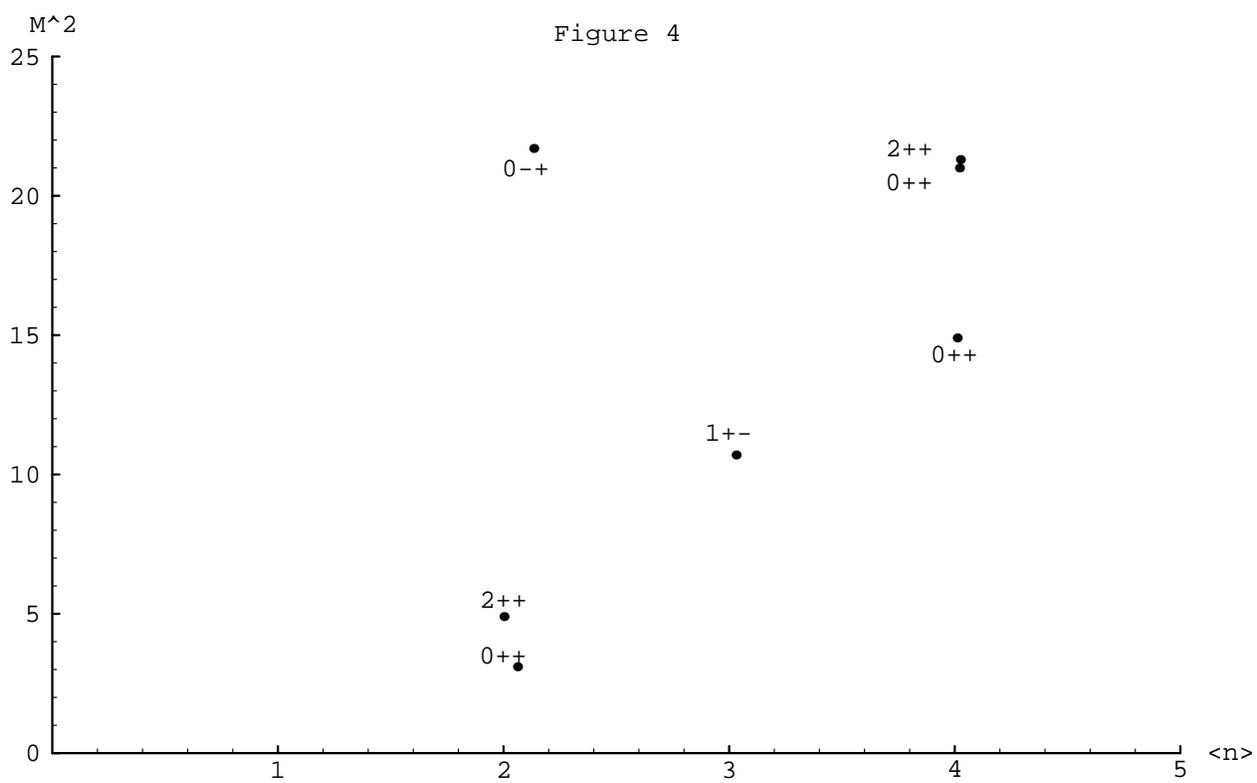
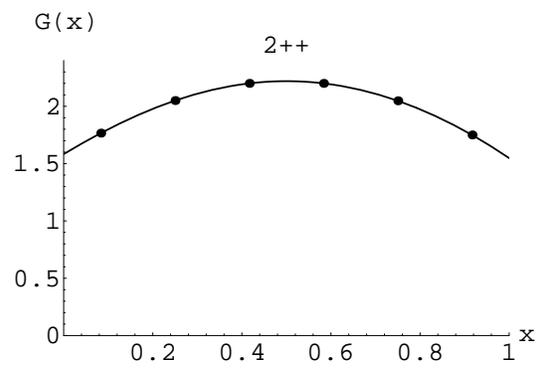
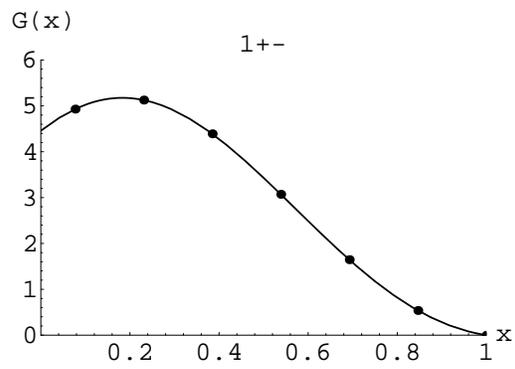
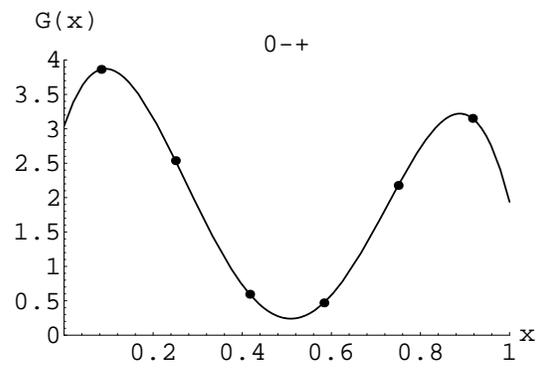
\end{document}